# Doubling the Critical Current Density of High Temperature Superconducting Coated Conductors through Proton Irradiation

Y. Jia,[1,#] M. LeRoux,[1] D. J. Miller,[1] J. G. Wen,[1] W. K. Kwok,[1] U. Welp,[1] M. W. Rupich,[2] X. Li,[2] S. Sathyamurthy,[2] S. Fleshler,[2] A. P. Malozemoff,[2] A. Kayani,[3] O. Ayala-Valenzuela,[4] L. Civale[4]

[1]Argonne National Laboratory, Argonne, IL 60439

[2]American Superconductor Corp., Devens, MA 01434

[3]Western Michigan University, Kalamazoo, MI 49008

[4]Los Alamos National Laboratory, Los Alamos, NM 87545

The in-field critical current of commercial $YBa_2Cu_3O_7$ coated conductors can be substantially enhanced by post-fabrication irradiation with 4 MeV protons. Irradiation to a fluence of $8 \times 10^{16}$ p/cm$^2$ induces a near doubling of the critical current in fields of 6 T || c at a temperature of 27 K, a field and temperature range of interest for applications such as rotating machinery. A mixed pinning landscape of preexisting precipitates and twin boundaries and small, finely dispersed irradiation induced defects may account for the improved vortex pinning in high magnetic fields. Our data indicate that there is significant head-room for further enhancements.

Tremendous improvements have been made during the past decade in the performance of YBa$_2$Cu$_3$O$_7$ (YBCO) coated conductors (CCs) for use in electric grid and magnet technologies [1, 2]. Production-line wires of hundreds of meters in length with critical current densities, $J_c$, exceeding 3 MA/cm$^2$ at 77 K and self-field are now being routinely manufactured. These wires have been successfully deployed in various demonstration projects world-wide [1]. Nevertheless, the rapid suppression of $J_c$, in magnetic fields, particularly applied along the *c*-axis, remains a barrier for the application of CCs in motors, transformers, generators, solenoids and MRI systems. For these applications, operation in magnetic fields of several Tesla and at temperatures around 30 K is envisioned [3]. Thus, improving the in-field performance of CCs is a priority in coated-conductor development.

Here, we demonstrate that irradiation of production-line CCs with 4-MeV protons to a fluence of 8x10$^{16}$ p/cm$^2$ induces a near doubling of the critical current $I_c = J_c\, t\, w$ (where *t* and *w* are the thickness and width of the YBCO film) in fields up to 6 T || c at a temperature of 27 K over the already high $I_c$-values of these conductors. We achieve at 27 K critical currents of 938 A/cm-w ($J_c$ = 7.8 MA/cm$^2$) and 634 A/cm-w ($J_c$ = 5.3 MA/cm$^2$) in applied fields of 3 T ||c and 6 T || c, respectively, performance levels useful for the applications mentioned above. We attribute the improved in-field performance to a mixed pinning landscape composed of preexisting precipitates and twin boundaries and small, finely dispersed irradiation induced defects. The improvements are obtained on post-fabrication CCs without the need to modify the precursor chemistry or film growth conditions. Our data indicate that there is significant head-room for even further enhancements.

The samples studied here are Dy-doped YBCO films of 1.2 μm nominal thickness deposited by MOD (metal-organic deposition) onto RABiTS™-substrates (Rolling Assisted Bi-axially Textured Substrates) [4]. Transport measurements were performed using a conventional four-terminal geometry on 2.5x3.0 mm² laser-cut bridges at 27 K (liquid Neon). $I_c$ was determined using the 1-μV/cm criterion. The magnetization was measured in a commercial SQUID magnetometer on 3x5 mm² rectangles patterned using photolithography and Ar-ion milling, and $J_c$ was determined using the Bean critical state model [5]. The flux creep rates were obtained from the time decay of the magnetization over a period of 1 hour. All samples included a 3 μm thick Ag-layer protecting the HTS film. Characterization of the defect structures was carried out using high-resolution and diffraction contrast transmission electron microscopy (TEM). TEM samples were prepared by focused ion beam (FIB) lift-out methods followed by back-side low-energy Ar ion milling. This approach minimizes artifacts due to specimen preparation. Several TEM samples were prepared from different regions of each sample (irradiated and unirradiated control) to ensure that results were representative of each condition. The samples were irradiated with 4 MeV-protons along the normal of the film plane (c-axis) at the 6-MV tandem van de Graaff accelerator at Western Michigan University. Typical p-beam currents of 200-300 nA were used. Proton irradiation has been used in numerous studies to enhance the vortex pinning in superconductors [6, 7], although typically not in production-line CCs [8]. Earlier TEM work on YBCO single crystals showed that p-irradiation induces a mixture of defects consisting of point defects (mostly on the O and Cu sites), and clusters of point defects and small collision cascades, both anisotropic and 2 – 5 nm in size [9].

The inset of Fig. 1 shows the evolution of the superconducting transition temperature $T_c$ with irradiation fluence. $T_c$ is suppressed by only ~ 1.5 K after a relatively high irradiation fluence of ~$8 \times 10^{16}$ p/cm$^2$. In comparison, previous work on YBCO single crystals revealed a suppression of $T_c$ by about 2 K after irradiation to $2 \times 10^{16}$ p/cm$^2$ with 3.5 MeV protons [6]. The main panel of Fig. 1 displays a comparison of the electric field versus current (*E-I*) curves at 27 K and 3 T || c for the reference sample and a sample irradiated to a fluence of $8 \times 10^{16}$ p/cm$^2$. The almost parallel shift of the curves reveals a substantial increase of $I_c$, in this case, from 143.6 A to 234.5 A. At the same time, the *E-I* curves hardly broaden as evidenced by the only slight decrease of the *n*-value (in $E \sim I^n$) from 28 for the reference sample to 25 for the irradiated sample. The p-irradiation induced enhancement of $I_c$ is also seen in the magnetization hysteresis data shown in Fig. 2 before and after irradiation to a fluence of $8 \times 10^{16}$ p/cm$^2$. The largest $I_c$ enhancement is observed at high fields where it approaches a factor of 1.8; the enhancement is substantially smaller at low fields (see below). A similar trend was also observed in recent low-energy Au-ion irradiation of YBCO films grown on SrTiO$_3$ single crystal substrates using a MOD process [10].

The evolution of the critical current and its enhancement with irradiation fluence and applied magnetic field is summarized in Fig. 3. Upon irradiation, $I_c$ is systematically enhanced, reaching 634 A/cm-w in a field of 6 T || c at *27 K*. For comparison, the 6T-value of a pristine wire at 27 K is only ~350 A/cm-w, and does

not reach 600 A/cm-w at temperatures above ~12K. A log-log plot of the data (Fig. 3a) reveals that in the field range of 1 T to 6 T $I_c$ for all p-irradiation fluences is well described by the relation $I_c \sim B^{-\alpha}$. Our data show that the major effect of p-irradiation is to diminish the magnetic-field induced suppression of $I_c$. Irradiation decreases the exponent $\alpha$, as shown in the inset for the transport and magnetization data, from ~0.75 for the reference sample to ~0.52 after irradiation to a fluence of $8 \times 10^{16}$ p/cm$^2$. Correspondingly, the beneficial effect of p-irradiation is larger at high field as shown directly in Fig. 3b. The $I_c$- enhancement obtained from the magnetization data follows the same trend as the transport data. The values are slightly lower for the same irradiation fluence due to the fact that the magnetization was measured at slightly higher temperature and that persistent current measurements use a significantly lower voltage criterion [5]. At low fields, deviations from the power-law relation occur as the enhancement tends to unity. In the low-field regime self-field effects and/or a change in pinning mechanism such as strong single-vortex pinning may become important. In fact, magnetization hysteresis data reveal that at fields below ~0.3 T, $I_c$ approaches a field-independent value (not shown). The steadily increasing $I_c$ with proton fluence in Fig. 3 also indicates that there is significant head-room for further enhancements with an even higher proton irradiation fluence. This is in contrast to results on single crystals for which, depending on the measurement temperature, irradiation fluences of $(1 – 5) \times 10^{16}$ p/cm$^2$ were found to be optimal [11]. Reported values of the exponent $\alpha$ depend strongly on the sample architecture and microstructure; for undoped PLD-grown films $\alpha$ - values of ~0.5 – 0.6 are common, whereas for MOD-grown films $\alpha$ - values are typically in the range of 0.6-0.7 [13], the latter being in good agreement with the results obtained here for the un-irradiated samples. Here we find that the introduction of a large number of small random defects via irradiation causes a significant reduction in $\alpha$. Previous reports on YBCO-films [14, 15] have shown that at high temperatures the exponent $\alpha$ decreases as more correlated defects are introduced into the pinning landscape. However for T<40K the power-law dependence was less well-defined, in contrast to our observation.

Figs. 4a, b show TEM images of the pristine and irradiated ($2 \times 10^{16}$ p/cm$^2$) sample with scattering vector (200), and Figs. 4 c, d show similar images with scattering vector (002). In the latter, the c-axis lattice spacing of YBCO is clearly visible. The pre-existing defect structure is characterized by rare-earth oxide precipitates several tens of nanometers in diameter [12]. Twin boundaries, which are also a common defect in this type of material, are not seen under the imaging conditions in Fig. 4; however, it has been observed that the precipitates tend to congregate on twin boundaries [12]. Proton-irradiation creates a fairly large number of small defects (2 – 5 nm) (Figs 4 b, d). Their different appearance under different imaging conditions indicates an anisotropic, most likely plate-like, shape. This type of defect is reproducibly seen in all imaged sections of the sample, and is consistent with similar findings on YBCO single crystals [9]. However, more work is required to establish the exact nature of the defects. Although a precise knowledge of the local thickness of the specimen is required to establish the density of the defects, an order of magnitude estimate may be obtained from previous results on single crystals which yielded a concentration of $\sim 2 \times 10^{16}$ cm$^{-3}$ following irradiation

with 3.5 MeV protons to a fluence of $2 \times 10^{16}$ p/cm$^2$ [9]. It should be noted though that the defect formation may strongly depend on the pre-existing microstructure, which is very complex for these highly optimized production CCs.

Guided by the images in Fig. 4 we propose a model of a mixed pinning landscape, which has proven very effective in optimizing vortex pinning. An example is the suppression of the motion of vortex kinks in a system of parallel columnar defects interspersed with point defects [14, 16]. Here we envision a model composed of large, strong pin-sites in form of rare-earth oxide precipitates and possibly twin boundaries in addition to a large number of weak, small irradiation-induced defects (see the top inset in Fig. 2) [17]. For the reference sample at low fields all vortices are assumed strongly pinned and $I_c$ is approximately field-independent. With increasing field $I_c$ is expected to initially vary approximately as $B^{-1/2}$ and, when interstitial vortices appear, as $B^{-1}$ [18]. Our data roughly follow this trend although $\alpha$ does not reach the value of -1. The low-field critical current does not significantly increase upon p-irradiation since all vortices are already strongly pinned (see Fig. 2). However, with increasing field the finely dispersed irradiation induced defects may effectively pin interstitial vortices (see Fig. 2). While this model can account for the general trends seen in our data, it remains to be seen whether the systematic evolution of the exponent $\alpha$ can be reproduced, for example in large-scale simulations as envisioned in the OSCon-SciDAC program [19].

The lower inset of Fig. 2 shows the temperature dependence of the normalized logarithmic flux creep rate, *S = dln(M)/dln(t)*, at 2T || c for various

irradiation fluences (where *M* is the magnetization). The qualitative features of the *S(T)* curves are similar to previous observations in YBCO, both single crystals and thin films [6, 11, 13, 21-23]. The initial increase of *S(T)* corresponds to an Anderson-Kim like creep with $S \approx T/U$, where U is the activation energy (approximately T-independent at low T), except that the nonzero extrapolation to *S(T=0)* is usually attributed to a quantum creep component. The unirradiated reference sample shows a peak in *S(T)* with a maximum at $T \sim 15\text{-}20K$, which becomes less pronounced after irradiation. This peak is characteristic of YBCO samples with disorder correlated along the *c*-axis [11, 13, 22] such as columnar or planar defects and presumably twin boundaries in our MOD films, and is due to the expansion of vortex double-kinks [20]. The temperature range where *S* is approximately constant has been identified earlier as a fingerprint of glassy relaxation [21], with values around $S \sim 0.02$ to $0.03$ being characteristic of collective creep of vortex bundles. Our data show a small but clear increase in *S(T)* as a function of irradiation fluence, indicating some level of sensitivity of the relaxation dynamics to the details of the pinning landscape. Recent studies [13, 23] have shown that unirradiated MOD films have lower *S*, particularly in the intermediate-temperature plateau, than other YBCO samples such as single crystals and films grown by pulsed laser deposition (PLD), an effect attributed to the larger pinning energy of the nanoparticles [13, 23]. At first glance, the increased flux creep with irradiation seems counterintuitive since one might expect more pinning to reduce the normalized creep rate *S*. However, the small increase in *S(T)* upon irradiation could be explained by the assumption that in the unirradiated sample pinning at intermediate temperatures and fields in the range of few Tesla is mostly due to nanoparticles while irradiation adds point and small cluster defects with a smaller characteristic glassy exponent $\mu$ and hence a higher plateau $S(T) \sim 1/\mu$. Finally, we note that the small increase in $S$ is also consistent with the slight decrease in the *n*-value based on the approximate relation *S ~ 1/(n-1)*, valid when the broadening of the *E-I* curves is mainly due to flux creep [24].

In summary, we have shown that irradiation of production-line CCs with 4 MeV-protons to a fluence of $8 \times 10^{16}$ p/cm² induces a near doubling of the critical current in fields of 6 T || c and at a temperature of 27 K over the already high $I_c$-values of these conductors. We achieve at 27 K critical currents of 938 A/cm-w ($J_c$ = 7.8 MA/cm²) and 634 A/cm-w ($J_c$ = 5.3 MA/cm²) in applied fields of 3 T ||c and 6 T || c, respectively. These values are very attractive for applications in motors, generators, transformers and magnet systems. A mixed pinning landscape composed of preexisting precipitates and twin boundaries and small, finely dispersed irradiation induced defects may account for the improved vortex pinning in high magnetic fields and may serve as a template for the design of effective high-field pinning structures. Our data indicate that there is significant head-room for further enhancements of $I_c$. These improvements are obtained on standard production wires without the need to modify the precursor chemistry or film growth conditions. Future work is needed to assess whether current accelerator technology can be adapted for a commercial roll-to-roll manufacturing process.

Acknowledgements: This work was supported by the Center for Emergent Superconductivity, an Energy Frontier Research Center funded by the U.S.

Department of Energy, Office of Science, Office of Basic Energy Sciences (YJ, ML, WKK, UW, OAV, LC) and by the Department of Energy, Office of Basic Energy Sciences, under Contract No. DE-AC02-06CH11357 (DJM, JGW). Irradiation of the samples was carried out at the Western Michigan University accelerator laboratory. Microstructural characterization was carried out in the Electron Microscopy Center at Argonne, which is supported by the Office of Science – Basic Energy Sciences.

Figure captions

Fig. 1. Current-electric field curves at 27 K and 3 T || c of a reference sample and a sample irradiated to a fluence of 8 x $10^{16}$ p/cm$^2$. The critical current, measured on a 2.5 mm wide bridge, is defined through the electric field criterion of 1 µV/cm. The inset shows the evolution of $T_c$ with irradiation fluence as determined from the magnetic transition measured in 10 Oe || c after zero-field cooling. Open and closed symbols represent pre and post-irradiation states, respectively.

Fig. 2. Magnetization hysteresis at 30 K before and after irradiation to 8 x $10^{16}$ p/cm2. The asymmetry of the magnetization hysteresis loops with respect to the field axis is a consequence of the ferromagnetic nature of the RABiTS–substrate. The lower inset shows the logarithmic relaxation rate *dln(M)/dln(t)* of the magnetization in a field of 2 T || c as function of temperature for various irradiation fluences. The upper inset shows a schematic of vortices in a mixed pinning landscape consisting of large strong pins (in blue) and numerous small pins (in red).

Fig. 3. Top: Field dependence of the critical current after various irradiation fluences on log-log-scales. $I_c$ is well described by $B^{-\alpha}$. The evolution of the exponent $\alpha$ with irradiation fluence is shown in the inset. Values of $\alpha$ obtained from magnetization measurements are included.
Bottom: Enhancement factor of the critical current as function of magnetic field and irradiation fluence.

Fig. 4. Diffraction contrast TEM images of pristine (a, c) and irradiated ($2\times10^{16}$ p/cm$^2$) (b, d) samples for diffraction vectors (200) (a, b) and (002) (c, d).

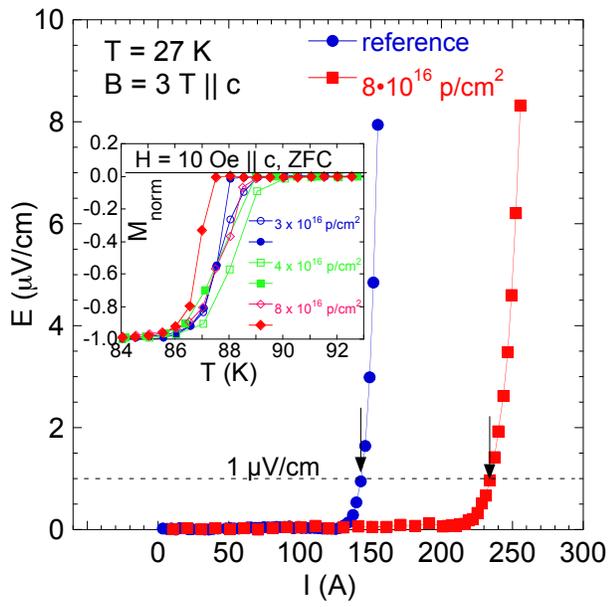

Fig. 1

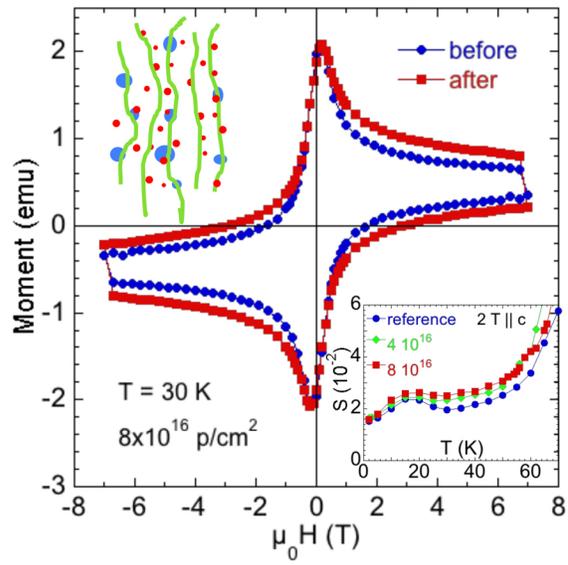

Fig. 2

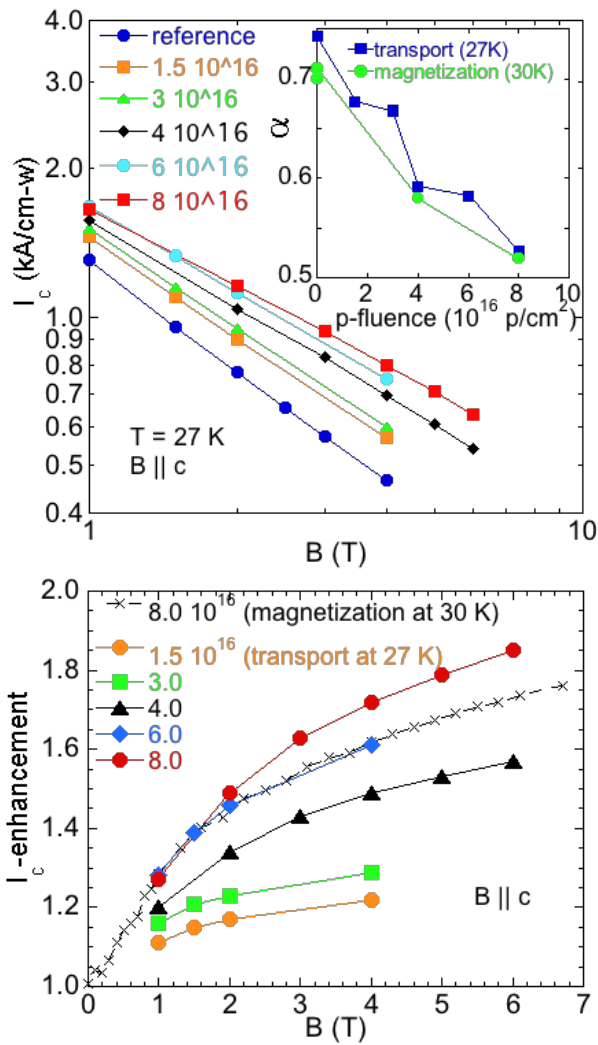

Fig. 3

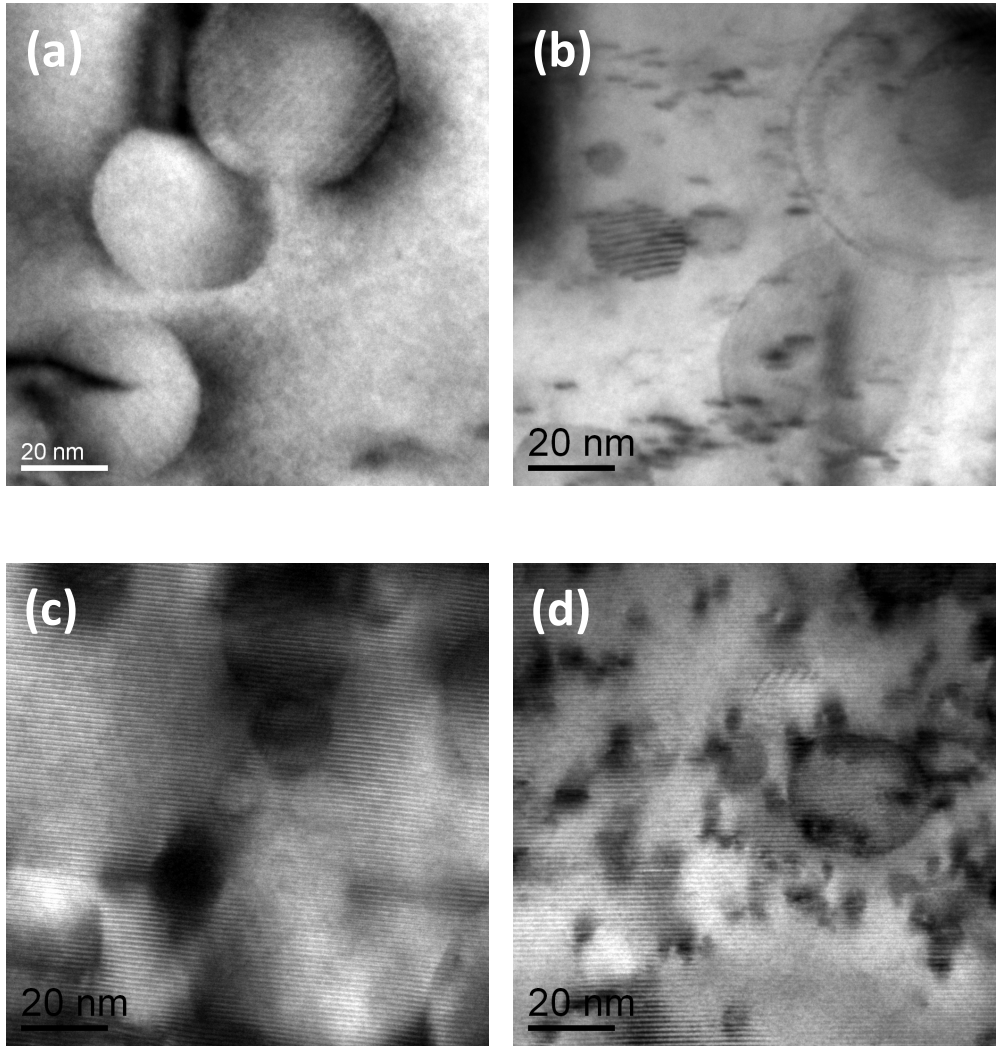

Fig. 4